%% file: main.tex
\documentclass[10pt,twocolumn,letterpaper]{article}

\usepackage{iccv}              %

\input{packages}

\input{macros}

\definecolor{iccvblue}{rgb}{0.21,0.49,0.74}
\usepackage[pagebackref,breaklinks,colorlinks,allcolors=iccvblue]{hyperref}
\usepackage[accsupp]{axessibility}  %

\def\paperID{***} %
\def\confName{ICCV}
\def\confYear{2025}

\makeatletter
\newcommand{\printfnsymbol}[1]{%
    \textsuperscript{\@fnsymbol{#1}}%
}
\makeatother

\newcommand\rurl[1]{%
  \href{https://#1}{\nolinkurl{#1}}%
}

\title{\model: Dynamic 3D Scene Generation from a Single Image and Actions}

\author{
Zizhang Li\textsuperscript{1}\printfnsymbol{1} \; 
Hong-Xing Yu\textsuperscript{1}\printfnsymbol{1} \;
Wei Liu\textsuperscript{1} \;
\\
Yin Yang\textsuperscript{2} \;
Charles Herrmann\textsuperscript{1} \;
Gordon Wetzstein\textsuperscript{1} \;
Jiajun Wu\textsuperscript{1} \;
\\[0.5em]
\textsuperscript{1}Stanford University \quad
\textsuperscript{2}University of Utah
\\
\\
{\url{https://kyleleey.github.io/WonderPlay/}}
}

\begin{document}

\input{shortcuts}

\twocolumn[\maketitle\input{figTexts/teaser}\bigbreak]

\def\thefootnote{*}\footnotetext[1]{Equal contribution}\def\thefootnote{\arabic{footnote}}

\input{texts/0_abstract}    
\input{texts/1_introduction}

\input{texts/2_related}

\input{texts/3_method}
\input{texts/4_experiments}
\input{texts/5_conclusion}

{
    \small
    \bibliographystyle{ieeenat_fullname}
    \bibliography{main}
}

\clearpage
\renewcommand\thefigure{S\arabic{figure}}
\setcounter{figure}{0}
\renewcommand\thetable{S\arabic{table}}
\setcounter{table}{0}
\renewcommand\theequation{S\arabic{equation}}
\setcounter{equation}{0}
\pagenumbering{arabic}%
\renewcommand*{\thepage}{S\arabic{page}}
\setcounter{footnote}{0}
\setcounter{page}{1}
\maketitlesupplementary
\appendix

\input{supp_text/results}

\input{supp_text/details}

\end{document}

%% file: packages.tex
\usepackage{graphicx}

%% file: macros.tex
\newcommand{\model}{WonderPlay\xspace}

\newcommand{\myparagraph}[1]{\vspace{0.1cm}\noindent\textbf{#1}}
\renewcommand{\paragraph}[1]{\vspace{0.1cm}\noindent\textbf{#1}}

\definecolor{MyDarkBlue}{rgb}{0,0.08,1}
\definecolor{MyAqua}{rgb}{0,0.7,0.7}
\definecolor{MyDarkGreen}{rgb}{0.02,0.6,0.02}
\definecolor{MyDarkRed}{rgb}{0.8,0.02,0.02}
\definecolor{MyDarkOrange}{rgb}{0.40,0.2,0.02}
\definecolor{MyPurple}{RGB}{111,0,255}
\definecolor{MyRed}{rgb}{1.0,0.0,0.0}
\definecolor{MyGold}{rgb}{0.75,0.6,0.12}
\definecolor{MyDarkgray}{rgb}{0.66, 0.66, 0.66}

\definecolor{Cardinal}{rgb}{0.549,0.082,0.082}

\newcommand{\zz}[1]{\textcolor{teal}{[Zizhang: #1]}}

\newcommand{\myicongravity}{%
  \raisebox{-0.5ex}{\includegraphics[height=1.2em]{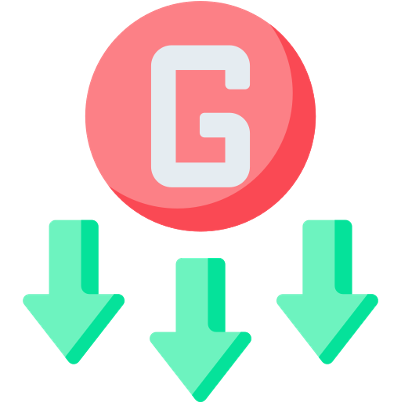}}%
}

\newcommand{\myiconforce}{%
  \raisebox{-0.5ex}{\includegraphics[height=1.2em]{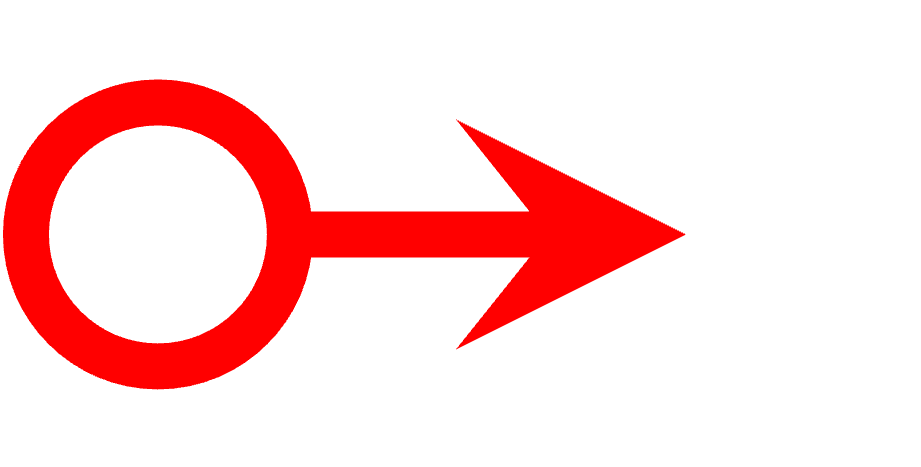}}%
}

\newcommand{\myiconwind}{%
  \raisebox{-0.5ex}{\includegraphics[height=1.2em]{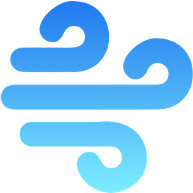}}%
}

%% file: shortcuts.tex
\newcommand{\ba}{\boldsymbol{a}}\newcommand{\bA}{\boldsymbol{A}}
\newcommand{\bb}{\boldsymbol{b}}\newcommand{\bB}{\boldsymbol{B}}
\newcommand{\bc}{\boldsymbol{c}}\newcommand{\bC}{\boldsymbol{C}}
\newcommand{\bd}{\boldsymbol{d}}\newcommand{\bD}{\boldsymbol{D}}
\newcommand{\be}{\boldsymbol{e}}\newcommand{\bE}{\boldsymbol{E}}
\newcommand{\bff}{\boldsymbol{f}}\newcommand{\bF}{\boldsymbol{F}} %
\newcommand{\bg}{\boldsymbol{g}}\newcommand{\bG}{\boldsymbol{G}}
\newcommand{\bh}{\boldsymbol{h}}\newcommand{\bH}{\boldsymbol{H}}
\newcommand{\bi}{\boldsymbol{i}}\newcommand{\bI}{\boldsymbol{I}}
\newcommand{\bj}{\boldsymbol{j}}\newcommand{\bJ}{\boldsymbol{J}}
\newcommand{\bk}{\boldsymbol{k}}\newcommand{\bK}{\boldsymbol{K}}
\newcommand{\bl}{\boldsymbol{l}}\newcommand{\bL}{\boldsymbol{L}}
\newcommand{\bm}{\boldsymbol{m}}\newcommand{\bM}{\boldsymbol{M}}
\newcommand{\bn}{\boldsymbol{n}}\newcommand{\bN}{\boldsymbol{N}}
\newcommand{\bo}{\boldsymbol{o}}\newcommand{\bO}{\boldsymbol{O}}
\newcommand{\bp}{\boldsymbol{p}}\newcommand{\bP}{\boldsymbol{P}}
\newcommand{\bq}{\boldsymbol{q}}\newcommand{\bQ}{\boldsymbol{Q}}
\newcommand{\br}{\boldsymbol{r}}\newcommand{\bR}{\boldsymbol{R}}
\newcommand{\bs}{\boldsymbol{s}}\newcommand{\bS}{\boldsymbol{S}}
\newcommand{\bt}{\boldsymbol{t}}\newcommand{\bT}{\boldsymbol{T}}
\newcommand{\bu}{\boldsymbol{u}}\newcommand{\bU}{\boldsymbol{U}}
\newcommand{\bv}{\boldsymbol{v}}\newcommand{\bV}{\boldsymbol{V}}
\newcommand{\bw}{\boldsymbol{w}}\newcommand{\bW}{\boldsymbol{W}}
\newcommand{\bx}{\boldsymbol{x}}\newcommand{\bX}{\boldsymbol{X}}
\newcommand{\by}{\boldsymbol{y}}\newcommand{\bY}{\boldsymbol{Y}}
\newcommand{\bz}{\boldsymbol{z}}\newcommand{\bZ}{\boldsymbol{Z}}

\newcommand{\balpha}{\boldsymbol{\alpha}}\newcommand{\bAlpha}{\boldsymbol{\Alpha}}
\newcommand{\bbeta}{\boldsymbol{\beta}}\newcommand{\bBeta}{\boldsymbol{\Beta}}
\newcommand{\bgamma}{\boldsymbol{\gamma}}\newcommand{\bGamma}{\boldsymbol{\Gamma}}
\newcommand{\bdelta}{\boldsymbol{\delta}}\newcommand{\bDelta}{\boldsymbol{\Delta}}
\newcommand{\bepsilon}{\boldsymbol{\epsilon}}\newcommand{\bEpsilon}{\boldsymbol{\Epsilon}}
\newcommand{\bzeta}{\boldsymbol{\zeta}}\newcommand{\bZeta}{\boldsymbol{\Zeta}}
\newcommand{\beeta}{\boldsymbol{\eta}}\newcommand{\bEta}{\boldsymbol{\Eta}} %
\newcommand{\btheta}{\boldsymbol{\theta}}\newcommand{\bTheta}{\boldsymbol{\Theta}}
\newcommand{\biota}{\boldsymbol{\iota}}\newcommand{\bIota}{\boldsymbol{\Iota}}
\newcommand{\bkappa}{\boldsymbol{\kappa}}\newcommand{\bKappa}{\boldsymbol{\Kappa}}
\newcommand{\blambda}{\boldsymbol{\lambda}}\newcommand{\bLambda}{\boldsymbol{\Lambda}}
\newcommand{\bmu}{\boldsymbol{\mu}}\newcommand{\bMu}{\boldsymbol{\Mu}}
\newcommand{\bnu}{\boldsymbol{\nu}}\newcommand{\bNu}{\boldsymbol{\Nu}}
\newcommand{\bxi}{\boldsymbol{\xi}}\newcommand{\bXi}{\boldsymbol{\Xi}}
\newcommand{\bomikron}{\boldsymbol{\omikron}}\newcommand{\bOmikron}{\boldsymbol{\Omikron}}
\newcommand{\bpi}{\boldsymbol{\pi}}\newcommand{\bPi}{\boldsymbol{\Pi}}
\newcommand{\brho}{\boldsymbol{\rho}}\newcommand{\bRho}{\boldsymbol{\Rho}}
\newcommand{\bsigma}{\boldsymbol{\sigma}}\newcommand{\bSigma}{\boldsymbol{\Sigma}}
\newcommand{\btau}{\boldsymbol{\tau}}\newcommand{\bTau}{\boldsymbol{\Tau}}
\newcommand{\bypsilon}{\boldsymbol{\ypsilon}}\newcommand{\bYpsilon}{\boldsymbol{\Ypsilon}}
\newcommand{\bphi}{\boldsymbol{\phi}}\newcommand{\bPhi}{\boldsymbol{\Phi}}
\newcommand{\bchi}{\boldsymbol{\chi}}\newcommand{\bChi}{\boldsymbol{\Chi}}
\newcommand{\bpsi}{\boldsymbol{\psi}}\newcommand{\bPsi}{\boldsymbol{\Psi}}
\newcommand{\bomega}{\boldsymbol{\omega}}\newcommand{\bOmega}{\boldsymbol{\Omega}}

\newcommand{\nA}{\mathbb{A}}
\newcommand{\nB}{\mathbb{B}}
\newcommand{\nC}{\mathbb{C}}
\newcommand{\nD}{\mathbb{D}}
\newcommand{\nE}{\mathbb{E}}
\newcommand{\nF}{\mathbb{F}}
\newcommand{\nG}{\mathbb{G}}
\newcommand{\nH}{\mathbb{H}}
\newcommand{\nI}{\mathbb{I}}
\newcommand{\nJ}{\mathbb{J}}
\newcommand{\nK}{\mathbb{K}}
\newcommand{\nL}{\mathbb{L}}
\newcommand{\nM}{\mathbb{M}}
\newcommand{\nN}{\mathbb{N}}
\newcommand{\nO}{\mathbb{O}}
\newcommand{\nP}{\mathbb{P}}
\newcommand{\nQ}{\mathbb{Q}}
\newcommand{\nR}{\mathbb{R}}
\newcommand{\nS}{\mathbb{S}}
\newcommand{\nT}{\mathbb{T}}
\newcommand{\nU}{\mathbb{U}}
\newcommand{\nV}{\mathbb{V}}
\newcommand{\nW}{\mathbb{W}}
\newcommand{\nX}{\mathbb{X}}
\newcommand{\nY}{\mathbb{Y}}
\newcommand{\nZ}{\mathbb{Z}}

\newcommand{\cA}{\mathcal{A}}
\newcommand{\cB}{\mathcal{B}}
\newcommand{\cC}{\mathcal{C}}
\newcommand{\cD}{\mathcal{D}}
\newcommand{\cE}{\mathcal{E}}
\newcommand{\cF}{\mathcal{F}}
\newcommand{\cG}{\mathcal{G}}
\newcommand{\cH}{\mathcal{H}}
\newcommand{\cI}{\mathcal{I}}
\newcommand{\cJ}{\mathcal{J}}
\newcommand{\cK}{\mathcal{K}}
\newcommand{\cL}{\mathcal{L}}
\newcommand{\cM}{\mathcal{M}}
\newcommand{\cN}{\mathcal{N}}
\newcommand{\cO}{\mathcal{O}}
\newcommand{\cP}{\mathcal{P}}
\newcommand{\cQ}{\mathcal{Q}}
\newcommand{\cR}{\mathcal{R}}
\newcommand{\cS}{\mathcal{S}}
\newcommand{\cT}{\mathcal{T}}
\newcommand{\cU}{\mathcal{U}}
\newcommand{\cV}{\mathcal{V}}
\newcommand{\cW}{\mathcal{W}}
\newcommand{\cX}{\mathcal{X}}
\newcommand{\cY}{\mathcal{Y}}
\newcommand{\cZ}{\mathcal{Z}}

\newcommand{\figref}[1]{Fig.~\ref{#1}}
\newcommand{\secref}[1]{Section~\ref{#1}}
\newcommand{\algref}[1]{Algorithm~\ref{#1}}
\newcommand{\eqnref}[1]{Eq.~\eqref{#1}}
\newcommand{\tabref}[1]{Table~\ref{#1}}

\def\mc{\mathcal}
\def\mb{\boldsymbol}

\newcommand{\T}{^{\raisemath{-1pt}{\mathsf{T}}}}

\newcommand{\Perp}{\perp\!\!\! \perp}

\makeatletter
\DeclareRobustCommand\onedot{\futurelet\@let@token\@onedot}
\def\@onedot{\ifx\@let@token.\else.\null\fi\xspace}
\def\eg{e.g\onedot,\xspace} \def\Eg{E.g\onedot,\xspace}
\def\ie{i.e\onedot,\xspace} \def\Ie{I.e\onedot,\xspace}
\def\cf{cf\onedot} \def\Cf{Cf\onedot}
\def\etc{etc\onedot}
\def\vs{vs\onedot}
\def\wrt{wrt\onedot}
\def\dof{d.o.f\onedot}
\def\etal{et~al\onedot}
\def\iid{i.i.d\onedot}
\def\evs{\emph{vs}\onedot}
\makeatother

\newcommand*\rot{\rotatebox{90}}

\newcommand{\boldparagraph}[1]{\vspace{0.4em}\noindent{\bf #1:}}

\definecolor{darkgreen}{rgb}{0,0.7,0}
\definecolor{lightred}{rgb}{1.,0.5,0.5}

\crefname{section}{Sec.}{Secs.}
\Crefname{section}{Section}{Sections}
\Crefname{table}{Table}{Tables}
\crefname{table}{Tab.}{Tabs.}

\newcommand{\method}[0]{WonderPlay\xspace}
\newcommand{\todo}[1]{\textcolor{red}{[TODO\@: #1]}}
\renewcommand{\zz}[1]{\textcolor{teal}{[ZZ\@: #1]}}
\newcommand{\jw}[1]{\textcolor{PineGreen}{[JW\@: #1]}}
\newcommand{\cih}[1]{\textcolor{magenta}{[CIH\@: #1]}}

\renewcommand{\paragraph}{%
  \@startsection{paragraph}{4}%
  {\z@}{-0.5em}{-0.5em}%
  {\normalfont\normalsize\bfseries}%
}

\newcommand{\ignore}[1]{}{}

%% file: figTexts/teaser.tex
\begin{center}
    \includegraphics[trim={0px 0px 0px 0px}, clip, width=\linewidth]{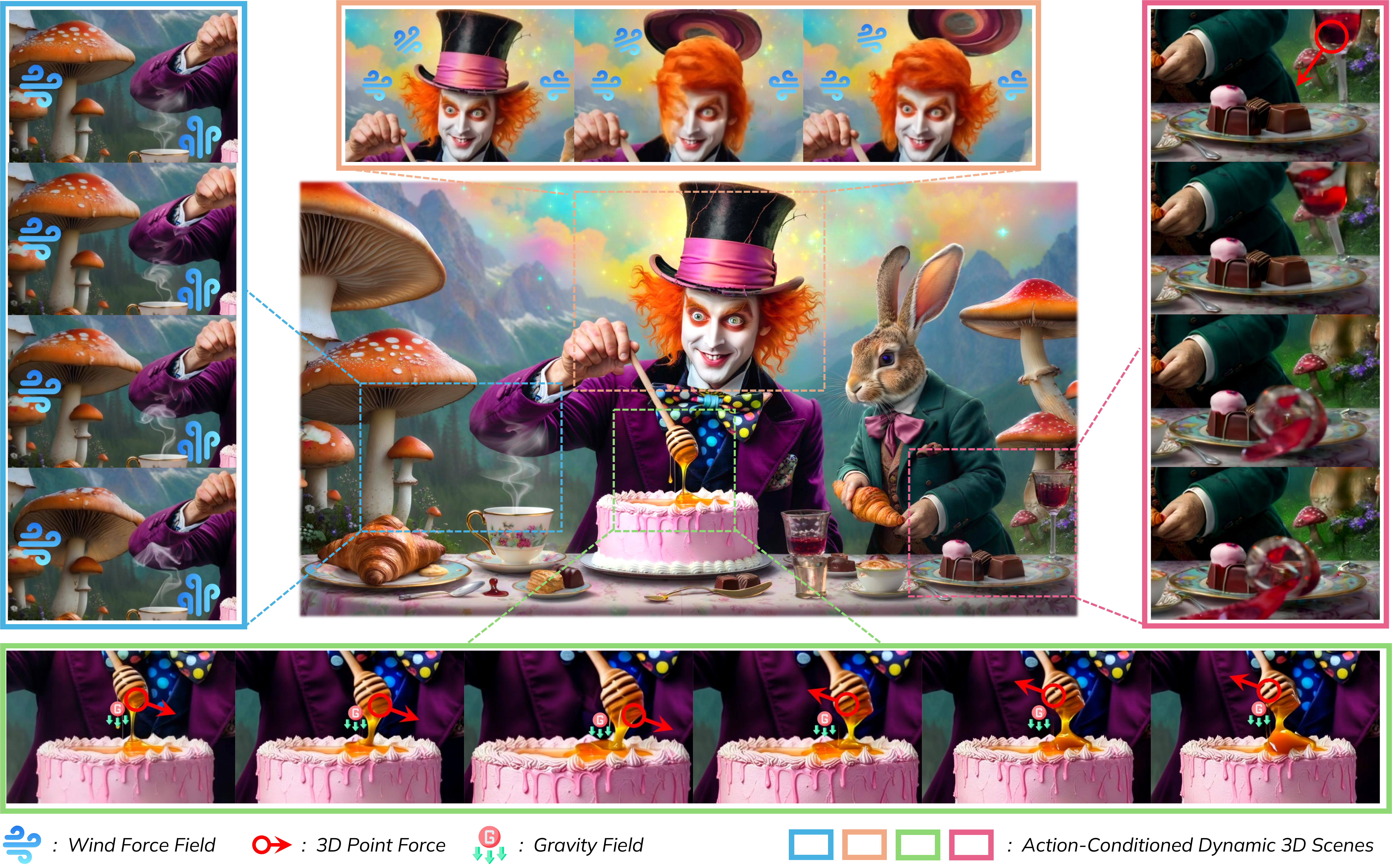}
\end{center}
\vspace{-0.2cm}
\captionof{figure}{
We propose \textbf{\method}, a framework that takes a single image and \textbf{actions} as inputs, and then generates dynamic 3D scenes that depict the consequence of the actions. \method allows users to interact with various scenes of diverse physical materials, \eg the hat and wine glass (rigid body), the hair (thin strands), the steam (gas), the mushroom (elastic), honey (liquid), and more. 
See {\url{https://kyleleey.github.io/WonderPlay/}} for interactive video results.
}%
\label{fig:teaser}

%% file: texts/0_abstract.tex
\begin{abstract}
\model is a novel framework integrating physics simulation with video generation for generating action-conditioned dynamic 3D scenes from a single image. While prior works are restricted to rigid body or simple elastic dynamics, \model features a hybrid generative simulator to synthesize a wide range of 3D dynamics. The hybrid generative simulator first uses a physics solver to simulate coarse 3D dynamics, which subsequently conditions a video generator to produce a video with finer, more realistic motion. The generated video is then used to update the simulated dynamic 3D scene, closing the loop between the physics solver and the video generator. This approach enables intuitive user control to be combined with the accurate dynamics of physics-based simulators and the expressivity of diffusion-based video generators. Experimental results demonstrate that \model enables users to interact with various scenes of diverse content, including cloth, sand, snow, liquid, smoke, elastic, and rigid bodies -- all using a single image input. 
Code will be made public. %

\vspace{-0.5em}

\end{abstract}

%% file: texts/1_introduction.tex
\vspace*{-8pt}\section{Introduction}
\label{s:intro}

Recent years have seen rapid progress in image, video, and 3D and 4D scene generation, culminating in models that achieve great visual quality~\cite{baldridge2024imagen,midjourney} and dynamic realism~\cite{sora2024}. This has directly motivated recent interest in generative world models, which, beyond their relevance to AR/VR and embodied AI~\citep{qureshi2024splatsim}, can also be created and explored as standalone experiences. However, while significant efforts have been devoted to enhancing the generation quality~\citep{bruce2024genie,che2024gamegen}, relatively little attention has been paid to enabling action-based interaction.
In this work, we %
study \textbf{action-conditioned dynamic 3D scene generation from a single image}: given an input image and a 3D action, such as wind or a point force, we aim to generate the resulting dynamic 3D scene in the near future. In particular, we focus on three types of actions: gravity, force fields like wind, and point forces like pushes or pulls.

Existing methods often rely exclusively on physics simulation for computing dynamic 3D scenes given user action input~\citep{xie2024physgaussian,zhang2024physdreamer,tan2024physmotion}. These methods face two critical limitations. First, they require \emph{accurate physics solvers for all types of dynamics} involved in the scene. Nevertheless, accurate physics solvers such as solid-fluid two-way coupling~\citep{chen2024solidfluid} still remain an open problem. Second, they require \emph{full reconstruction of physical states from limited observations}. However, reconstructing complete physical states for materials like snow, sand, cloth, and fluids from a single image is often infeasible. Consequently, existing methods are constrained to a narrow range of dynamics types, primarily rigid body dynamics~\citep{liu2024physgen} and simple elasticity~\citep{zhang2024physdreamer,tan2024physmotion}.

This motivates us to incorporate data priors from video generation models~\citep{sora2024,bar2024lumiere,yang2024cogvideox}, which are trained on extensive real-world videos of diverse physical phenomena. However, video generation models cannot accept precise 3D actions as inputs and simulate the resulting dynamics. In this work, we reconsider the relationship between physics simulators and video generation models for action-conditioned dynamic 3D scene generation. Our novel framework, \emph{\textbf{\model}}, enables users to interact with 3D scenes encompassing diverse materials—including rigid bodies, cloth, liquids, gases, and granular substances—from a single input image, as shown in Figure~\ref{fig:teaser}.

Our core technical idea is a \textbf{hybrid generative simulator}. First, we let the physics simulator provide a coarse simulation of action-induced dynamic consequences to a video generator. Conditioned on the coarse simulation, the video generator synthesizes a video with realistic motion. Finally, the synthesized video is used to update the coarse simulation.

In the conditional video generation stage, we explore a novel strategy to optimally use the simulator conditioning signal: a motion--appearance \emph{bimodal control} scheme, designed to improve the quality and realism of the dynamics in the generated video. Additionally, to reduce the video generator hallucination in simulator-trustable spatial regions such as static backgrounds, we  introduce a \emph{spatially varying} masking scheme for the bimodal control.

In summary, our contributions are three-fold: 
\begin{itemize}
    \item We tackle the challenging problem of single-image, action-conditioned dynamic 3D scene generation with diverse physical materials.
    \item We propose \model, featuring a hybrid generative simulator that integrates a physics solver and video diffusion to acquire both high simulation fidelity in response to actions and high visual quality.
    \item We demonstrate that \model significantly outperforms both pure physics-based methods and adapted video generation models in terms of visual quality and physical plausibility under various interactions.
\end{itemize}

%% file: texts/2_related.tex
\section{Related Work}%
\input{figTexts/approach}

\myparagraph{Action-conditioned dynamic scene generation.} Early work on generating action-conditioned dynamic scenes approached the problem by extracting modal bases of vibrating objects in 2D image space~\citep{davis2015visual,davis2015image}, essentially representing motion as a series of vibrations with different frequencies and intensities. Following the advent of generative diffusion modeling~\cite{sohldickstein2015diffusion,ho2020denoising,song2020denoising}, this approach was later extended by retaining the same motion representation but generating the modal basis with a diffusion model~\citep{li2024generative}. While this representation can be effective for motions similar to vibrations, modal basis functions struggle to represent more general motions, prompting the emergence of an alternative line of research that explicitly uses physics solvers. For example, PhysGen~\citep{liu2024physgen} focused on the 2D domain using a rigid-body physics solver to handle colliding objects.

Recently, several physics-based approaches have been developed to synthesize dynamic 3D scenes~\citep{chen2022virtual,zhong2024reconstruction,le2023differentiable,xie2024physgaussian,zhang2024physdreamer,huang2024dreamphysics,liu2024physics3d, lin2024phys4dgen, aira2024motioncraft,chen2025physgen3d}. However, due to the requirements of physics solvers, all of these techniques require complete 3D geometric reconstructions of the scene,  requiring complex, multi-view captures. For example, Virtual Elastic Objects~\citep{chen2022virtual} reconstructs the geometry, appearances, and physical parameters of elastic objects from a multi-view capture setup. Later work, such as PAC-NeRF~\citep{li2023pac}, PhysGaussian~\citep{xie2024physgaussian}, and PhysDreamer~\citep{zhang2024physdreamer}, integrates physics-based simulations with NeRF or 3D Gaussians from multi-view reconstruction.
A concurrent approach, PhysMotion~\citep{tan2024physmotion}, is closest to our work. Both approaches take a single image as input and use a combination of a 3D physics solver and a video generation model. Unlike ours, however, PhysMotion~\citep{tan2024physmotion} relies on a physics solver to compute the dynamics for the entire scene, only using the video generator to refine the appearance. Due to the restrictive assumption that the physics solver will specify all the dynamics, PhysMotion is limited to rigid and elastic dynamics. In contrast, WonderPlay uses both a physics solver and the video generator to compute dynamics, enabling realistic action-conditioned dynamics for various types of physical phenomena, and generates a dynamic 3D scene as opposed to a video.

\myparagraph{Controllable video generation.} In recent years, video generation has rapidly improved, making significant strides in both visual quality and realistic dynamics~\citep{makeavideo2022,imagenvideo2022,svdblattmann2023,emuvideo2023,bar2024lumiere,sora2024,yang2024cogvideox}. Recent video generation methods, such as Sora~\citep{sora2024}, have demonstrated great promise in generating diverse real-world physical phenomena. However, despite this promise, these models are conditioned with text and/or images and lack controllability regarding general actions and other physical inputs. While there has been considerable work on adding controls to video models, most of this work has focused on camera control~\citep{he2024cameractrl,zheng2024cami2v,sun2024dimensionx,xiao2024video,hou2024training,xu2024camco} and various types of motion control~\citep{yin2023dragnuwa, chen2023mcdiff, wang2024motionctrl,shi2024motion,niu2024mofa,wu2024draganything,geng2024motion,li2024image,namekata2024sg, zhang2024tora}, including drag-based, trajectory-based, and optical flow-based approaches. 
However, most of these motion control models require the resulting dynamics of an action as input to generate a trajectory-following video. Concurrently, Motion Prompting~\citep{geng2024motion} uses temporally sparse trajectories as conditioning to generate videos that adhere to an initial trajectory and then continue using a generative video prior. Nevertheless, many actions, such as those involving fluids or wind, are difficult or impossible to represent as trajectory signals. In contrast, we aim to take physics-based 3D actions as input and model a dynamic 3D scene, rather than just a video.

\myparagraph{World models.}
Along with the rise in video models, there has been a growing interest in interactive world models~\citep{ha2018world}, which recurrently generate world states from prior state and actions. While this area has seen considerable research, the focus has been on video game domains~\cite{valevski2024diffusion, che2024gamegen, bruce2024genie} due to the data availability of action-video pairs. Consequently, both the generated worlds and the actions considered are centered around those found in video games. Only a few concurrent works have explored this for the real world (e.g.,~\cite{bar2024navigation, agarwal2025cosmos}) and none provides interaction beyond camera control or text. In contrast, we focus on physics-based actions in realistic worlds.

\myparagraph{Dynamic 3D scene generation.} 
Text-conditioned generation of 3D scenes with motion has primarily been tackled by distilling video generation into dynamic 3D representations~\citep{singer2023text}. Most of the existing works focus on objects or simple scenes composed of a few objects~\citep{zhao2023animate124,bahmani20244d,ling2024align,ren2023dreamgaussian4d,bah2024tc4d}, while recent methods also attempt to deal with more complex scenes including nature~\citep{lee2024vividdream,liu2024dynamicscaler}. Recent work~\citep{watson2024controlling, zhao2024genxd, wu2024cat4d} has focused on training generators which model 4D itself, conditioning on time and camera pose. Yet, they do not have the ability to simulate dynamics in response to actions.
\label{s:related}

%% file: figTexts/approach.tex
\begin{figure*}[t]
  \centering
   \includegraphics[width=\linewidth]{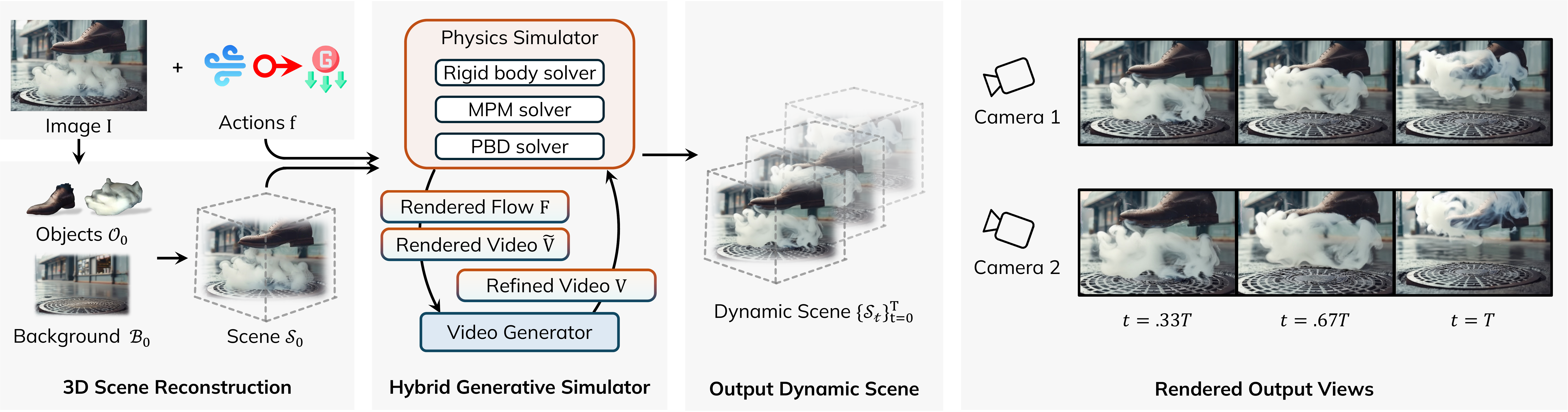}

   \caption{\textbf{Overview of \method.} Given a single image, we first reconstruct the 3D scene and estimate material properties. Then our hybrid generative simulator uses physics solver and input actions to infer coarse 3D dynamics. The simulated appearance and motion signals are used to condition the video generator through spatially varying bimodal control to synthesize the realistic motion. The dynamic 3D scene is refined using the synthesized video, finishing the hybrid generative simulation.}
   \vspace{-0.1cm}
   \label{fig:approach}
\end{figure*}

%% file: texts/3_method.tex
\vspace*{-4pt}\section{\model}

\myparagraph{Formulation.} Our goal is action-conditioned 3D scene dynamics synthesis. The input is a single image $\mathbf{I}$ and actions. We model actions as three types of forces: gravity $\mathbf{f}_\text{g}$, 3D force fields $\mathbf{f}_\text{w}(x,y,z,t)$ such as wind, and 3D point forces $\mathbf{f}_\text{p}(t)$ which is defined on a point of an object. The output is a dynamic 3D scene $\{\mathcal{S}_t\}_{t=0}^T$ that is the consequence of applying the actions to the input scene, where $\mathcal{S}_0$ denotes our initial 3D scene representation recovered from the input image $\mathbf{I}$, and $T$ denotes the total simulation time steps. 

\myparagraph{Overview.} We aim to simulate the dynamics of diverse materials including rigid, elastic, cloth, smoke, liquid, granular, and their interactions. To this end, we propose \model
. As illustrated in Figure~\ref{fig:approach}, we first reconstruct the 3D scene $\mathcal{S}_0$ from the input image $\mathbf{I}$ (top left of Figure~\ref{fig:approach}). Our main technical innovation is the hybrid generative simulator (middle left of Figure~\ref{fig:approach}). It takes the 3D scene $\mathcal{S}_0$ and the actions as input and predicts the 3D dynamics $\{\mathcal{S}_t\}_{t=1}^T$ (middle right of Figure~\ref{fig:approach}).

\subsection{3D Scene Reconstruction}

Our 3D scene representation $\mathcal{S}_t=\mathcal{B}_t\cup\mathcal{O}_t$ consists of a background $\mathcal{B}_t$ and objects $\mathcal{O}_t$ at a timestep $t$.
Our first step is to reconstruct/generate $\mathcal{S}_0$ from the input image $\mathbf{I}$. 
We reconstruct the 3D background $\mathcal{B}_0$ and the 3D objects $\mathcal{O}_0$ separately to jointly form $\mathcal{S}_0=\mathcal{B}_0\cup\mathcal{O}_0$.

\myparagraph{Background.} We represent the background with Fast Layered Gaussian Surfels (FLAGS)~\citep{yu2024wonderworld}.
Formally, the background $\mathcal{B}_t=\{\mathbf{p}^\text{B}, \mathbf{q}^\text{B}, \mathbf{s}^\text{B}, \mathbf{o}^\text{B}, \mathbf{c}_{t}^\text{B}\}$ consists of $N_\text{B}$ Gaussian surfels, parameterized by 3D spatial positions $\mathbf{p}^\text{B}\in\mathbb{R}^{3N_\text{B}}$, orientation quaternions $\mathbf{q}^\text{B}$, scales $\mathbf{s}^\text{B}$, opacities $\mathbf{o}^\text{B}$, and view-independent RGB colors $\mathbf{c}^\text{B}_t$. 
We detail the reconstruction of background in Appendix~\ref{supp:background} and treat it as a static boundary in simulation.

\myparagraph{Objects.}
An ``object'' in WonderPlay refers to a dynamic entity we simulate in the physics solver, including rigid object, cloth, granular material, and fluids.
To represent a simulatable object that is compatible with our physics solvers, we build a simulation-ready representation on top of the Gaussian surfels by adding connectivity to them, turning them into ``topological Gaussian surfels''. Formally, the topological Gaussian surfels consist of $N_\text{O}$ Gaussian surfels with edges and velocities, $\mathcal{O}_t=\{\mathbf{E}, \mathbf{v}_{t},\mathbf{p}_{t}^\text{O}, \mathbf{q}_{t}^\text{O}, \mathbf{s}_{t}^\text{O}, \mathbf{o}_{t}^\text{O}, \mathbf{c}_{t}^\text{O}\}$, where the edge matrix $\mathbf{E}\in\{0,1\}^{N_\text{O}\times N_\text{O}}$ indicates the topological connectivity of the surfels, and $\mathbf{v}_{t}\in\mathbb{R}^{3N_\text{O}}$ denotes the velocity.

We create the initial topological Gaussian surfels $\mathcal{O}_0$ by first generating an object mesh from an image segment of the object using an image-to-mesh model InstantMesh~\citep{xu2024instantmesh}. Then, we bind a Gaussian surfel to each of the mesh vertices. We detail this process in Appendix~\ref{app:topological}.

\myparagraph{Materials.}
Besides the geometry and appearance representation $\mathcal{O}$, an object also has material properties $\mathbf{m}$.
The definition of object material depends on the object type, which follows a 6-way classification: rigid, elastic, cloth, smoke, liquid, and granular. We detail the material properties in Appendix~\ref{app:material}.

\subsection{Hybrid Generative Simulator}

\myparagraph{Main idea.}
The reconstructed scene geometry $\mathcal{S}_0$ and estimated material properties $\mathbf{m}$ are inherently inaccurate and incomplete, and accurate physics solvers for all materials and their complex interactions are still an open problem. Therefore, existing methods are limited to simple rigid/elastic simulations~\citep{zhang2024physdreamer,liu2024physgen,tan2024physmotion}. Our main idea to address this challenge is extracting the dynamics knowledge from a video generator which has been trained on numerous videos of real-world physics. 

In particular, we use physics solvers to estimate a coarse and incomplete dynamic scene $\{\tilde{\mathcal{S}}_t\}_{t=1}^T$ given initial scene $\mathcal{S}_0$ and actions $\mathbf{f}_\text{g}$, $\mathbf{f}_\text{w}$, $\mathbf{f}_\text{p}$. The coarse dynamic scene is used to drive the video generator to synthesize a video $V$ that has realistic dynamics. We obtain the output dynamic scene $\{\mathcal{S}_t\}_{t=1}^T$ by updating $\{\tilde{\mathcal{S}}_t\}_{t=1}^T$ to match the video $V$ through differentiable rendering.

\myparagraph{Physics solvers.} At each simulation time step, a physics solver takes the current scene $\tilde{\mathcal{S}}_t$ and forces $\mathbf{f}_\text{g}$, $\mathbf{f}_\text{w}(t)$, $\mathbf{f}_\text{p}(t)$ as input, and solves for the object dynamics attributes including the velocity $\mathbf{v}_{t+1}$, position $\mathbf{p}_{t+1}^\text{O}$, and orientation $\mathbf{q}_{t+1}^\text{O}$ at the next time step:
\begin{equation}
    \mathbf{v}_{t+1},\mathbf{p}_{t+1}^\text{O}, \mathbf{q}_{t+1}^\text{O} = \texttt{solver}(\tilde{\mathcal{S}}_t, \mathbf{f}_\text{g}, \mathbf{f}_\text{w}(t),\mathbf{f}_\text{p}(t)),
\end{equation}
where $\tilde{\mathcal{S}}_0=\mathcal{S}_0$.
Then, we construct the coarse scene at the next time step $\tilde{\mathcal{S}}_{t+1}$ by
\begin{equation}
    \tilde{\mathcal{S}}_{t+1} = \mathcal{B}_0 \cup \{\mathbf{E}, \mathbf{v}_{t+1},\mathbf{p}_{t+1}^\text{O}, \mathbf{q}_{t+1}^\text{O}, \mathbf{s}_{0}^\text{O}, \mathbf{o}_{0}^\text{O}, \mathbf{c}_{0}^\text{O}\},
\end{equation}
where we keep all non-dynamics attributes the same as $\mathcal{S}_0$.
To compute the dynamics attributes of various materials, we employ multiple types of physics solvers. These solvers are coupled to tackle multi-physics scenes, e.g., fluid and rigid as shown in Figure~\ref{fig:approach}. Please see details for different solvers in Appendix~\ref{app:material}.

\input{figTexts/approach_video_model}

\input{figTexts/exp_comparison}

\myparagraph{Conditioning the video generator.} 
Given the coarse dynamic scene $\{\tilde{\mathcal{S}}_t\}_{t=0}^T$, we condition a video generator to synthesize a video $\mathbf{V}\in\mathbb{R}^{(T+1)\times H\times W\times 3}$ that has more detailed motion while adhering to the action consequence depicted by the coarse dynamic scene. To this end, we introduce a \emph{bimodal control} scheme that uses two modalities for control: motion (represented by flow) and appearance (RGB). In particular,
\begin{equation}
    \mathbf{V}=g(\mathbf{F},\tilde{\mathbf{V}},\mathbf{I}),
\end{equation}
where $g$ denotes the video generator, $\mathbf{F}\in\mathbb{R}^{T\times H\times W\times 2}$ denotes the pixel-space flow rendered using the velocity $\{\mathbf{v}_t\}_{t=1}^T$, $\tilde{\mathbf{V}}\in\mathbb{R}^{(T+1)\times H\times W\times 3}$ denotes the video rendered from the coarse scene $\{\tilde{\mathcal{S}}_t\}_{t=0}^T$, and $\mathbf{I}$ denotes the input image. We show an illustration in Figure~\ref{fig:approach-video-model}.

\vspace{0.1cm}
\noindent
\underline{Motion control.}
We leverage a pre-trained motion-controlled image-to-video diffusion model, Go-with-the-Flow~\citep{burgert2025go}, as our $g$. The motion control is based on noise warping. In short, instead of using an unstructured random Gaussian noise distribution, it uses a warping-based structured noise $\mathbf{N}(\mathbf{F})\in\mathbb{R}^{(T+1)\times H\times W\times 3}$. $\mathbf{N}(\mathbf{F})$ is created by first sampling a random Gaussian $\mathbf{N}_0\in\mathbb{R}^{H\times W\times 3}$ and then iteratively doing warping such that $\mathbf{N}_{t+1}=\texttt{warp}(\mathbf{N}_t,\mathbf{F}_{t+1})$ where $\mathbf{F}_{t+1}\in\mathbb{R}^{H\times W\times 2}$ denotes the flow at $t+1$. The structured noise $\mathbf{N}(\mathbf{F})$ is then fused with some random noise to improve visual quality, controlled by a degradation factor $\gamma$~\citep{burgert2025go}.\footnote{The video model is a latent diffusion model~\citep{yang2024cogvideox} which downsamples the spatiotemporal dimensions $H,W,T$ and upsamples feature dimension from $3$ to $C$. But for notational simplicity, we do not distinguish the latent space from the pixel space.}

\vspace{0.1cm}
\noindent
\underline{RGB control.}
To incorporate additional control with RGB frames $\tilde{\mathbf{V}}$, we use  SDEdit~\citep{meng2021sdedit}. Specifically, the diffusion-based generation process is gradually denoising $\mathbf{N}(\mathbf{F})$, such that $\mathbf{V}_{s-1}=\texttt{Denoise}(\mathbf{V}_s,s)$ where $s=S,S-1,\cdots,1$ denotes the diffusion timestep, $\mathbf{V}_S=\mathbf{N}(\mathbf{F})$ is the initial noise, and the generated video is given by $\mathbf{V}=\mathbf{V}_0$. We control this process by skipping first several steps and directly starting the denoising from step $s_1<S$ with
\begin{equation}
    \mathbf{V}_{s_1}=\alpha_{s_1}\tilde{\mathbf{V}}+\sqrt{1-\alpha_{s_1}^2}\mathbf{N}(\mathbf{F}),
\end{equation}\label{eqn:vs1}%
where $\alpha_i$ denotes the diffusion coefficient at timestep $i$. This has been shown to control the main content of the generation, while allowing details to be synthesized~\citep{meng2021sdedit}.

\vspace{0.1cm}
\noindent
\underline{Discussion.} With our bimodal control, we pass the coarse motion and appearance information from the physics simulator to the video generator. This allows not only generating more realistic motion, but also fixes appearance artifacts caused by imperfect 3D scene reconstruction and the lack of lighting information to resolve appearance changes. The question is how much we trust the generator to overwrite the coarse information. Intuitively, this is dictated by $s_1$: If $s_1$ is close to $0$, then the generator only modifies the coarse video $\tilde{\mathbf{V}}$ a bit; if $s_1$ is close to $S$, then it can overwrite $\tilde{\mathbf{V}}$ more and hallucinate new contents. Thus, $s_1$ positively corresponds to the responsibility of the video generator.

\myparagraph{Spatially varying responsibility.} 
The responsibility of the video generator is inherently uneven across spatial regions in every frame. For example, most of our background remains static in the dynamic process, which we want to trust the simulator output $\tilde{\mathbf{V}}$ more, rather than the video generator, because the video generator may hallucinate incorrect details such as ghost objects. To this end, we introduce the spatially varying bimodal control.

In this work, we consider two responsibility levels in our spatially varying bimodal control for the background and the dynamic objects, respectively, such that we set a lower responsibility $s_2<s_1$ of the video generator on modifying the background. Specifically, at the step $s_2$, we compute
\begin{equation}
    \hat{\mathbf{V}}_{s_2} = \mathbf{M}\odot\mathbf{V}_{s_2}+(\mathbf{1}-\mathbf{M})\odot(\alpha_{s_2}\tilde{\mathbf{V}}+\sqrt{1-\alpha_{s_2}^2}\mathbf{N}(\mathbf{F})),
\end{equation}
where $\mathbf{V}_{s_2}$ is computed from gradually denoising $\mathbf{V}_{s_1}$. $\mathbf{M}\in\{0,1\}^{(T+1)\times H\times W\times 3}$ denotes the binary mask that uses $1$ to mark a pixel of dynamic objects and $0$ to mark a pixel of the background, which is rendered from the coarse scene. After this computation, we set $\mathbf{V}_{s_2}\leftarrow \hat{\mathbf{V}}_{s_2}$ and continue the denoising to generate $\mathbf{V}$.

\myparagraph{Updating scene dynamics.} Finally, we use the generated video $\mathbf{V}$ as a supervision to update the coarse dynamic scene $\{\tilde{\mathcal{S}}_t\}_{t=0}^T$. This is done by minimizing a photometric L1 loss: $\min_{\{\mathbf{c}^\text{B}_t,\mathcal{O}_t\}_{t=0}^T} \lVert \mathbf{V}-\tilde{\mathbf{V}} \rVert_1$ over the foreground object's motion trajectory and appearance $\{\mathcal{O}_t\}_{t=0}^T$. We also update the background color $\mathbf{c}_t^\text{B}$ for shading effects. This optimization yields the final dynamic scene $\{\mathcal{S}_t\}_{t=0}^T$.

%% file: figTexts/approach_video_model.tex
\begin{figure}[t]
  \centering
   \includegraphics[width=0.8\linewidth]{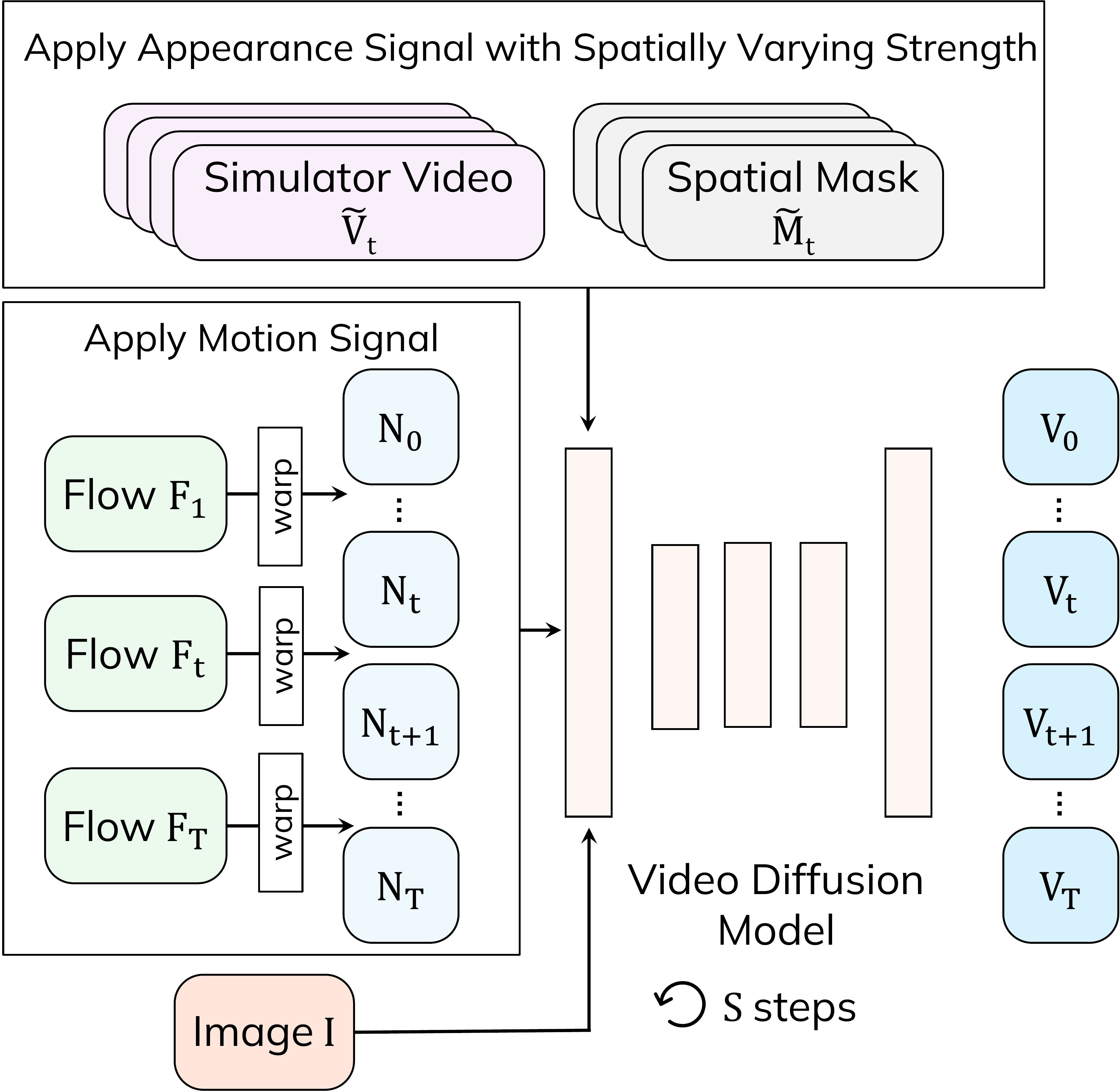}

   \caption{Illustration on our spatially varying bimodal control, which drives the video generator with input image $\mathbf{I}$, pixel-space flow $\mathbf{F}$ and simulation rendered $\tilde{\mathbf{V}}$.}
   \vspace{-0.1cm}
   \label{fig:approach-video-model}
\end{figure}

%% file: figTexts/exp_comparison.tex
\begin{figure*}[t]
    \centering
    \includegraphics[width=\linewidth]{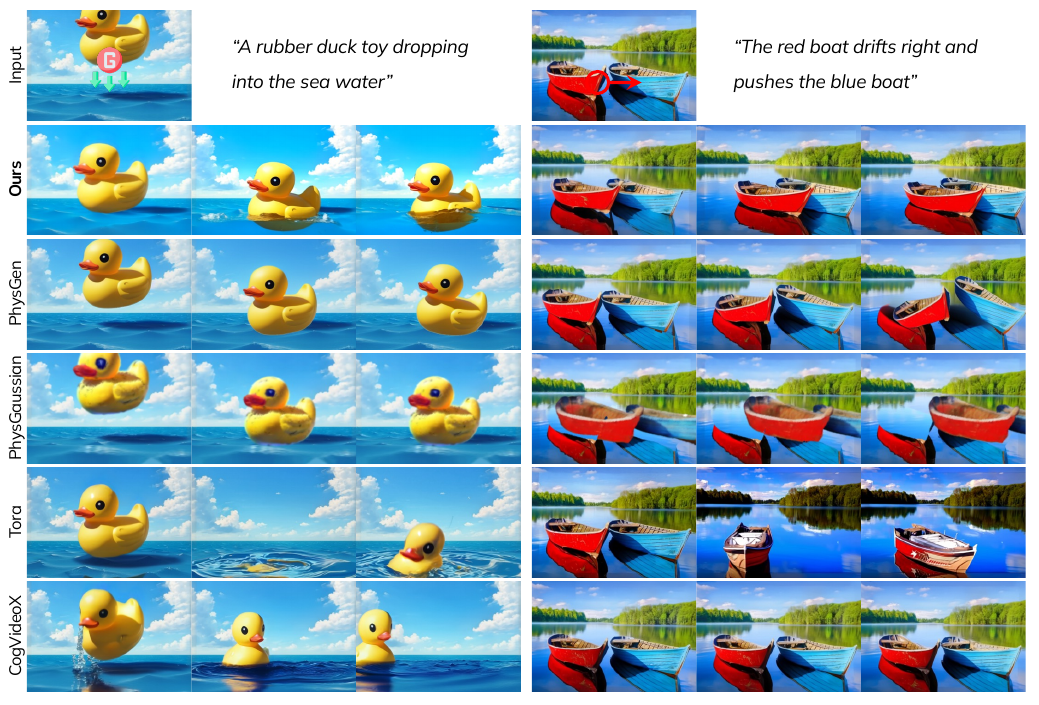}
    \caption{\textbf{Qualitative comparisons} between \method~(ours) and the baseline methods. The top row shows the input images, actions, and the texts describing the actions for CogVideoX~\citep{yang2024cogvideox}. 
    }
    \vspace*{-6pt}
    \label{fig:comparison}
\end{figure*}

%% file: texts/4_experiments.tex
\section{Experiments}%
\label{s:experiments}

\myparagraph{Implementation details.} 
For physics simulation, we adopt the Genesis~\cite{Genesis} framework, which unifies several different physics solvers. For all  scenes, we run physical simulation for $960$ steps and render one frame for each $20$ steps.
We include additional implementation details in Appendix~\ref{supp:impl}.

\myparagraph{Baselines.} We compare against two types of baselines for action-conditioned 3d dynamic scene generation: physics-based and conditional video generation methods.
For physics-based methods, we compare with PhysGen~\cite{liu2024physgen} and PhysGaussian~\cite{xie2024physgaussian}. PhysGen decomposes an image into 2D rigid bodies and run rigid simulation given certain action. PhysGaussian models the 3D scene as elastic objects with the MPM~\cite{jiang2016material} framework. Since PhysGen only requires a single image as the input, we directly follow their preprocessing for image decomposition. PhysGaussian requires multiview images to reconstruct the underlying scene first, so we provide as input our reconstructed 3D scene and then run simulation with given actions.
For conditional video generation methods, we compare against two methods: CogVideoX-I2V~\cite{yang2024cogvideox} with text prompts and Tora~\cite{zhang2024tora} with drag-based conditioning. For Tora, we use the trajectories from our simulation as the drag input.

\myparagraph{Metrics.} We render videos from the input viewpoint to compute quantitative metrics. We adopt the imaging, aesthetic, motion quality, and consistency metrics from VBench~\citep{huang2024vbench}. We also adopt the GPT-4o-based physical realism metric~\citep{chen2025physgen3d}. We curate 15 examples, including 7 real photos and 8 realistic synthetic images, covering diverse types of scenes contents including cloth, rigid body, elastic objects, liquid, gas, granular substance, etc.

\input{figTexts/exp_qualitative_results}
\input{tables/user_study}

\input{tables/quantitative}

\subsection{Results}

\input{figTexts/exp_diff_interactions}
\input{figTexts/exp_ablation_vid}
\input{figTexts/exp_ablation}

\myparagraph{Comparison to baselines.}
We show side-by-side comparisons on two scenes in Figure~\ref{fig:comparison}. The top row shows input images, actions~(gravity for duck dropping, force to pull the red boat towards the right) and the text prompt for CogVideoX-I2V~\cite{yang2024cogvideox}, followed by the action-conditioned generated dynamics from our method and the baselines. 

Despite their ability to produce plausible visual quality, video generation methods struggle to adhere to the actions. In the duck-dropping-into-water scene, Tora~\citep{zhang2024tora} submerges the duck under the water and then changes its shape after it re-emerges. CogVideoX-I2V struggles to generate realistic dynamics for the duck's drop and adds undesirable dynamics by moving the duck to the left. Both models also struggle with the boats scene. Tora completely alters the scene mid-video, while CogVideoX-I2V fails to generate meaningful dynamics.

As for the physics-based method, PhysGen~\cite{liu2024physgen} is limited to rigid body simulation in 2D space, making it hard to handle scene with complex materials such as water. PhysGaussian~\cite{liu2024physgen} typically requires multiview images and as a result, struggles to produce a reasonable 3D representation with only one input image. Also due to the lack of a complete 3D physical state, both physics-based methods fail to properly handle the shading effect in the boats scene and make the reflections move with the boats. 
Our method, in contrast, offers the advantages of both physical simulation and video generation: the physical simulation handles a wide range of materials and ensures the desired dynamics, and the video generation model provides visual realism by successfully synthesizing water waves and bubbles surrounding the duck, as well as the following reflections.
Also shown in Table~\ref{tab:quantitative},~\model~(ours) achieve the best or second-best performance across all metrics, showing strong motion quality, visual quality, and physical plausibility.

\myparagraph{User study.}
To evaluate the generation results with human preference, we recruit 200 participants and conduct a user study. We employ a Two-alternative Forced Choice (2AFC) protocol. Each participant evaluates 10 scenes. The participants view an action description alongside a randomly ordered side-by-side comparison video: one from our method and one from a baseline. Participants then select which video demonstrates superior performance in one of three criteria: \textit{physics plausibility} which measures the correctness of the predicted motion in response to the action, \textit{motion fidelity } that reflects the quality and naturalness of the generated motion, and \textit{visual quality}. 

We show the averaged results on all scenes in Table~\ref{tab:user-study}. 
In comparison to all baselines, about $70\%$ to $80\%$ of the participants prefer \method~(ours) across all three aspects, proving the superior performance of combining the physical simulator and video generator for dynamics with fidelity in response to actions and realistic visual appearance.

\myparagraph{Diverse scenes and materials.}
In Figure~\ref{fig:qualitative}, we present the generated dynamic 3D scenes on a variety of input images with diverse actions. It is important to note that achieving realistic visual quality in simulations of complex materials from a single image input with limited physical state information is extremely challenging. However, with the aid of the video generator, the sticky jam appears vivid as it pours onto the cake, and the river waves look natural in response to the boat's movement. Notably, the underlying physical simulator ensures that all dynamics follow the input actions. For example, the roses are initially blown to the right by the wind and then move back due to their elasticity.

\myparagraph{Condition on different actions.}
A significant advantage of our method is that it enables generating different interactions with different actions in the same scene. In Figure~\ref{fig:diff-interact}, we present four scenes, each with two different actions and their corresponding output dynamic 3D scenes. 

\subsection{Ablation on Hybrid Generative Simulator}
In the following we discuss an ablation study on the hybrid generative simulator. We leave quantitative numbers and further ablation in Appendix~\ref{sec:additional_exp}.

\myparagraph{Video generator refines both dynamics and appearance.}
In Figure~\ref{fig:ablation_vid}, we compare a dynamic scene created solely with the physics simulator, i.e., the coarse simulation (top row), and the refined dynamic scene created by our full model (bottom row). In the coarse simulation, we observe unrealistic motion: the motion of the smoke looks too sticky due to numerical viscosity that exists for almost all fluid solvers. The video model refines it so that the fluid motion looks smooth with swirls. There are also appearance artifacts in the coarse simulation such as the grainy smoke, where the video model can also refine them.

\myparagraph{Both signals to condition the video generator are necessary.}
To demonstrate the benefit of conditioning the video generator on both motion and appearance signals, we show ablation results in Figure~\ref{fig:ablation}. The top row shows the synthesized video from our full model with both signals; ``w/o RGB'' uses motion but no appearance; and ``w/o flow'' uses appearance but no motion. 
Using only RGB conditioning (``w/o flow''), the video model fails to retain or improve detailed dynamics in the sand grains. Using only a motion signal (``w/o RGB'') leads to unexpected hallucinations beyond user action input, \eg it hallucinates a pile of sand standing in the back and the background texture unexpectedly changes. In contrast, using both signals produces the best results.

%% file: figTexts/exp_qualitative_results.tex
\begin{figure*}
    \centering
    \includegraphics[width=\linewidth]{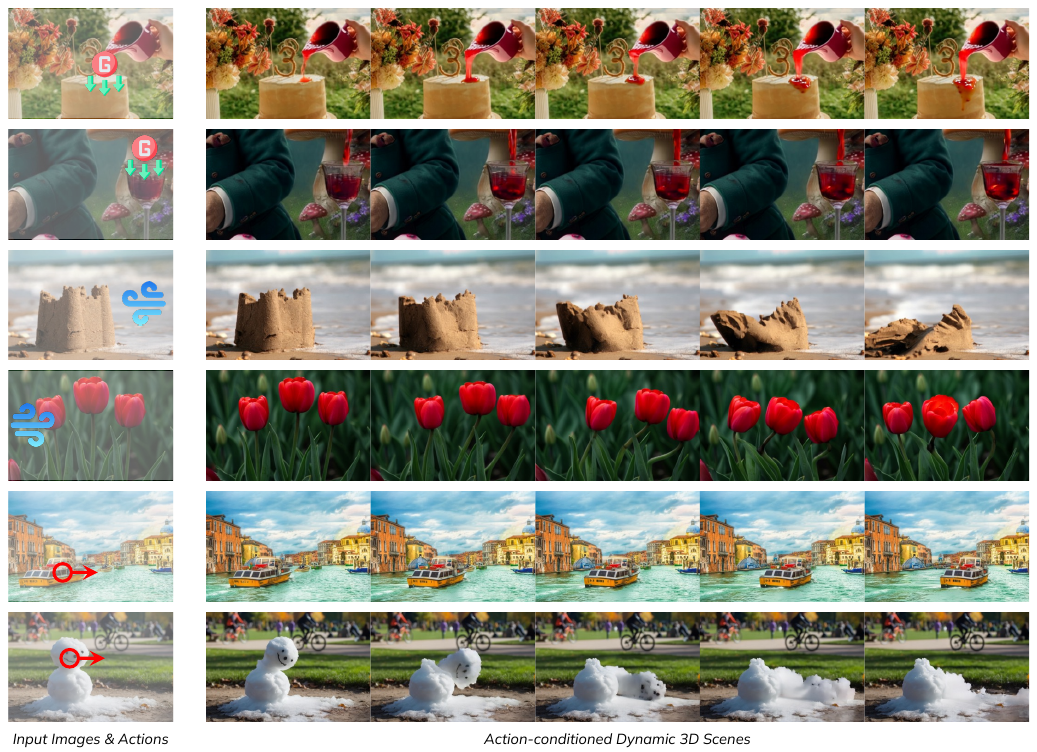}
    \caption{\textbf{Qualitative results} of the proposed \method. In the left column we show the input scene image and actions, where 
    \myicongravity{}, \myiconwind{}, \myiconforce{} 
    indicate gravity action, wind field action and 3D point force action, respectively. }
   \vspace{-5pt}
    \label{fig:qualitative}
\end{figure*}

%% file: tables/user_study.tex
\begin{table}[t]
    \scriptsize
	\setlength{\tabcolsep}{2pt} %
    \renewcommand{\arraystretch}{1.2}
	\centering
	\begin{tabular}{lcccc}
\toprule
                        & \textbf{Physics Plausibility} & \textbf{Motion Fidelity} & \textbf{Visual Quality} &  \\
\midrule
Over PhysGen~\citep{liu2024physgen}       & $78.0\%$         & $78.0\% $         & $80.1\%$           &  \\
Over PhysGaussian~\citep{xie2024physgaussian}  & $80.2\%$         & $81.2\%$          & $85.2\%$           &  \\
Over Tora~\citep{zhang2024tora}          & $77.0\%$         & $72.0\%$          & $71.0\% $          &  \\
Over CogVideoX-I2V~\citep{yang2024cogvideox} & $80.2\%$         & $73.0\% $         & $74.6\%$           &  \\
\bottomrule
\end{tabular}
	\caption{Human study 2AFC results of favor rate of \method~(Ours) over baseline methods.}
\label{tab:user-study}
\vspace{-0.1cm}
\end{table}

%% file: tables/quantitative.tex
\begin{table}[t]
\centering
\scriptsize
\setlength{\tabcolsep}{1pt}
\renewcommand{\arraystretch}{1.2}
\begin{tabular}{lccccc}
\toprule
Methods & Imaging ($\uparrow$) & Aesthetic ($\uparrow$) & Motion ($\uparrow$) & Consistency ($\uparrow$) & PhysReal ($\uparrow$) \\
\midrule
PhysGen & \underline{$0.692$} & $0.593$ & $0.992$ & $0.212$ & $0.545$ \\
PhysGaussian & $0.492$ & $0.564$ & \underline{$0.994$} & $0.206$ & $0.350$ \\
CogVideoX & $0.686$ & $0.574$ & $0.993$ & $\mathbf{0.219}$ & $0.670$ \\
Tora & $0.644$ & $\mathbf{0.620}$ & $0.992$ & $0.210$ & $0.530$ \\
\textbf{Ours} & $\mathbf{0.695}$ & \underline{$0.610$} & $\mathbf{0.995}$ & \underline{$0.217$} & $\mathbf{0.700}$ \\
\bottomrule
\end{tabular}
\caption{Quantitative comparison to baselines on 15 scenes.}
\vspace{-0.1cm}
\label{tab:quantitative}
\end{table}

%% file: figTexts/exp_diff_interactions.tex
\begin{figure*}[t]
    \centering
    \includegraphics[width=\linewidth]{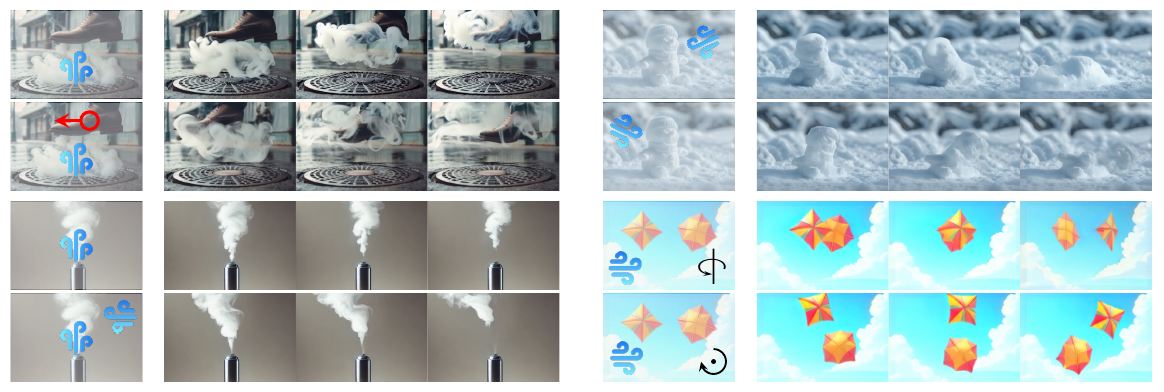}
    \caption{\textbf{Different actions} on the same scene. \method supports to use different 3D actions on the same scene. Here we show four different scenes and the corresponding dynamics from two different actions within each scene.}
    \label{fig:diff-interact}
    \vspace{-5pt}
    
\end{figure*}

%% file: figTexts/exp_ablation_vid.tex
\begin{figure}[t]
  \centering
   \includegraphics[width=\linewidth]{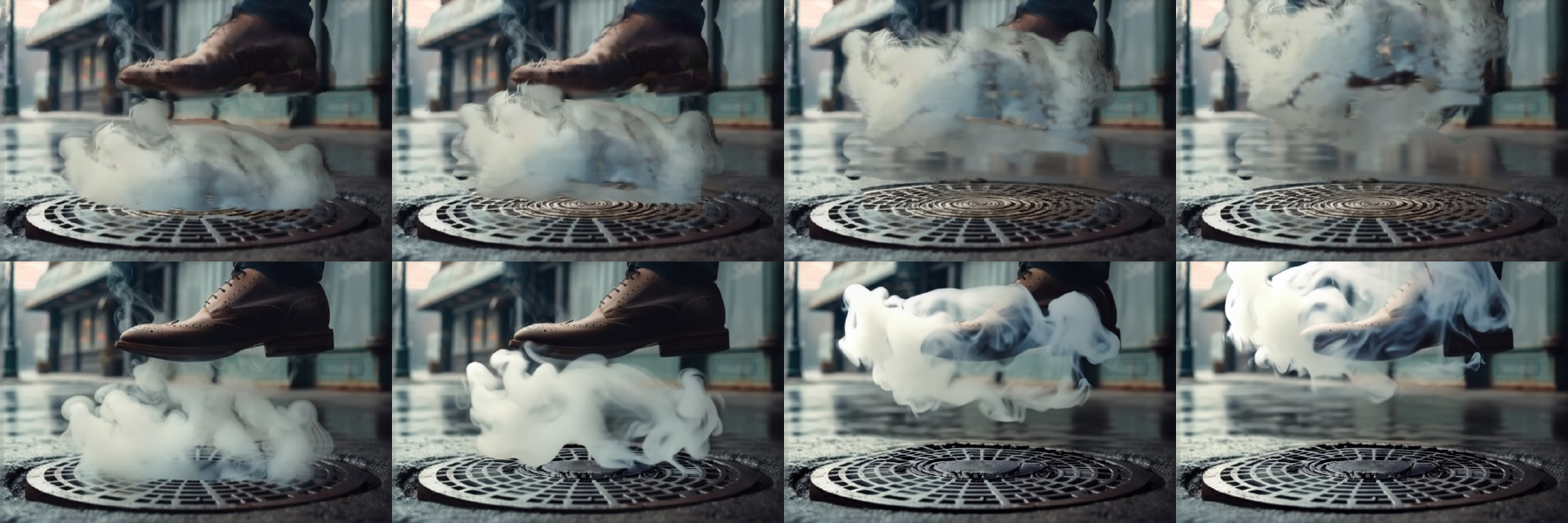}

   \caption{\textbf{Ablation} on hybrid generative simulator. Top row: Coarse simulation (i.e., only physics solver is used without video generator for refinement). Bottom row: Refined dynamic scene.}
   \label{fig:ablation_vid}
\end{figure}

%% file: figTexts/exp_ablation.tex
\begin{figure}[t]
  \centering
   \includegraphics[width=\linewidth]{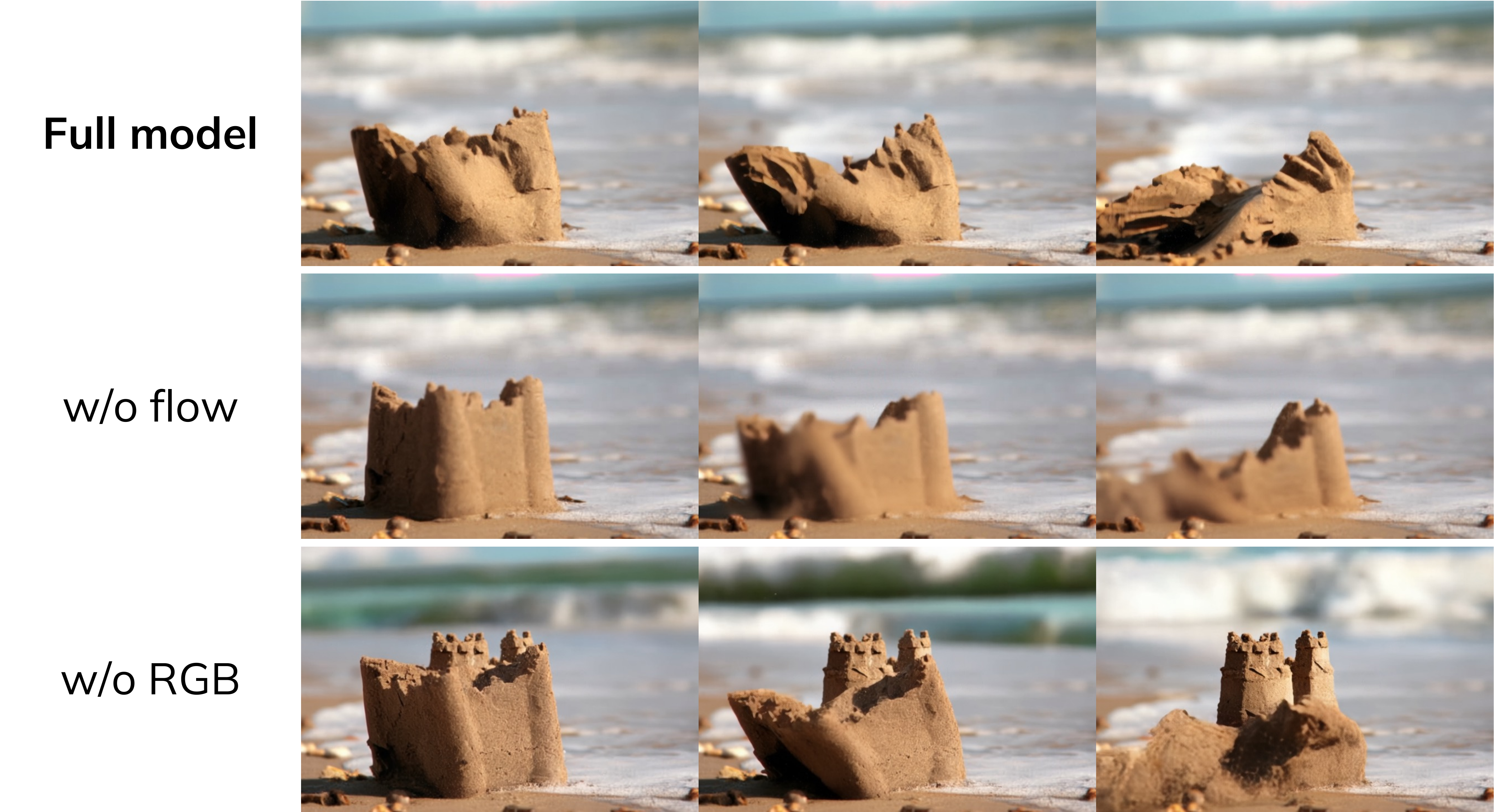}

   \caption{\textbf{Ablation} on the motion signal and the appearance signal to condition the video generator.}
   \label{fig:ablation}
\end{figure}

%% file: texts/5_conclusion.tex
\vspace{0.1cm}\section{Conclusion}
\label{s:conclusion}

In this work, we propose \method, a novel framework for action-conditioned dynamic 3D scene generation from a single image. \method features a hybrid generative simulator for simulation fidelity and visual quality. We showcase superior performance of \method on diverse scenes with various interactions.

\myparagraph{Acknowledgments.}
We thank Guandao Yang, Yunzhi Zhang, and Zhehao Li for the comments and fruitful discussions, and Hadi Alzayer for help reviewing the draft. This work is in part supported by the Stanford Institute for Human-Centered AI (HAI), the Okawa Foundation Research Grant, NSF RI \#2211258 and \#2338203, ONR YIP N00014-24-1-2117, and ONR MURI N00014-22-1-2740.

%% file: supp_text/results.tex
\section{Additional Evaluation}\label{sec:additional_exp}
\myparagraph{Additional ablation.}
We provide additional ablation on diffusion parameters including $s_1$, $s_2$, and $\gamma$ in Table~\ref{tab:additional_ablation}. The performances of different configurations slightly drop compared to our optimal values. In particular, lower $s_1$ and $s_2$ insufficiently leverage video priors, while higher values could weaken adherence to the physics simulation. Similarly, lower $\gamma$ under-leverage video priors.

\myparagraph{Staged evaluation.}
We design a staged evaluation with increasing complexity in a series of scenes (Figure~\ref{fig:staged_scenes}): a simple rigid ball falling onto a desk~(stage 1), make the ball elastoplastic~(stage 2), replace the desk with water to form multiphysics~(stage 3), and replace the ball with a duck for more complex shape~(stage 4). As shown in the diagrams in Figure~\ref{fig:staged_results}, while baseline methods perform well in early stages with simple physics, our method significantly outperforms baselines in later stages where the scenes involve complex physics and geometry.

%% file: supp_text/details.tex
\section{Technical Details}
\subsection{Additional Implementation Details}
\label{supp:impl}
When conditioning the video generator with simulated dynamics, we use the standard resolution and time values: $H=480$, $W=720$, with $T=48$ frames~(in total $49$ frames output). 
For sampling, we use a DDIM~\cite{song2020denoising} scheduler and iterate $S=25$ steps on the warped noise for the final video output.
We empirically set degradation factor $\gamma = 0.4$ and apply appearance signal at $s_1=21, s_2=18$ diffusion steps , as this combination usually provides the optimal results.

\subsection{Reconstructing Background}
\label{supp:background}
In a nutshell, the initial scene background $\mathcal{B}_0$ is generated by decomposing the input image $\mathbf{I}$ into several image layers, unprojecting all pixels in each layer to 3D space with estimated depth~\citep{ke2024repurposing}, followed by a photometric optimization to match the rendering with the input image $\mathbf{I}$ via differentiable rendering~\citep{kerbl20233d}. We refer the reader to~\citet{yu2024wonderworld} for more details of the generation process.

\subsection{Reconstructing Topological Gaussian Surfels}\label{app:topological}
To reconstruct the 3D objects by the topological Gaussian surfels from the input image $\mathbf{I}$, we first segment the object image by the Segment Anything Model~\citep{kirillov2023segment} and then we apply an image-to-mesh generation model InstantMesh~\citep{xu2024instantmesh}. In addition to the mesh, InstantMesh also generates multi-view object images $\{\mathbf{I}_i\}$ at fixed viewpoints as intermediate outputs. We bind a Gaussian surfel to each of the mesh vertices. Specifically, we first initializing a Gaussian surfel at a vertex with the vertex normal and the vertex color, and then we optimize the Gaussian surfel parameters so that the rendered images matches the multi-view images $\{\mathbf{I}_i\}$ via differentiable rendering~\citep{yu2024wonderworld,kerbl20233d}.

However, up to here the topological Gaussian surfels are still in a canonical coordinate frame. We need to register each of objects back to the scene coordinate frame. To do this, we first estimate the object orientation by DUSt3R~\citep{Wang_2024_CVPR}, and then we solve for a scale $s$ and a 3D translation $\mathbf{T}$ by least square to align the two coordinate frames. This requires us to find 3D correspondences to form the least square objective. We sample 3D points in the scene coordinate frame by first sampling pixels in $\mathbf{I}$ within the object segment, and then unprojecting the object pixels to 3D with the estimated depth~\citep{ke2024repurposing}, similar to the background. To sample 3D points in the object canonical frame, we sample object pixels from the image rendered from the object representation with the DUSt3R-estimated pose. Each of these pixels uniquely correspond to a 3D point in the object canonical frame.

For stabler simulation, we also adopt the internal filling technique as in PhysGaussian~\citep{xie2024physgaussian}.

\subsection{Material and Physics Solvers}\label{app:material}
We consider homogeneous uniform materials, i.e., $\mathbf{m}$ is constant within an object.
We follow~\citet{liu2024physgen} to estimate the values of the material parameters $\mathbf{m}$ by a Vision-Language Model (VLM) with optional manual adjustment for physical plausibility during simulation.

Here we provide further complementary information on each object material model and their solvers. 
We consider homogeneous uniform materials, i.e., $\mathbf{m}$ is constant within an object.
To model an object, we follow~\citet{liu2024physgen} to do a 6-way classification (rigid, elastic, cloth, smoke, liquid, and granular) by a VLM, and estimate the values of the material parameters $\mathbf{m}$ by the VLM with optional manual adjustment for physical plausibility during simulation. The material models and solvers are as follows.

\input{figTexts/staged_scenes}
\input{figTexts/staged_results}

\input{tables/additional_ablation}

\myparagraph{Rigid body.} 
We model a rigid object as a strictly undeformable mesh without internal links. The material properties $\mathbf{m}$ of a rigid object includes the density $\rho$ and the friction coefficient $k$. Recall that our topological Gaussian surfels are given by: $\mathcal{O}_t=\{\mathbf{E}, \mathbf{v}_{t},\mathbf{p}_{t}^\text{O}, \mathbf{q}_{t}^\text{O}, \mathbf{s}_{t}^\text{O}, \mathbf{o}_{t}^\text{O}, \mathbf{c}_{t}^\text{O}\}$, where the edge matrix $\mathbf{E}\in\{0,1\}^{N_\text{O}\times N_\text{O}}$ indicates the topological connectivity of the surfels, and $\mathbf{v}_{t}\in\mathbb{R}^{3N_\text{O}}$ denotes the velocity. They can be seen as a super-set of a mesh that has $\mathbf{E}$ and $\mathbf{p}$. Therefore, we can directly apply a rigid body solver to simulate our rigid objects.
We adopt a rigid body solver based on shape matching~\citep{muller2005meshless}. 
At each simulation time step, the rigid solver uses action forces to update the dynamics attributes, and then detects collisions among rigid objects and the background using connectivity information $\mathbf{E}$ to resolve penetrations.

\myparagraph{Elastic, liquid, and granular materials.}
We model these materials with continuum mechanics and simulate them using a Material-Point-Method (MPM) solver~\citep{jiang2016material}, similar to PhysGaussian~\citep{xie2024physgaussian}. The material properties $\mathbf{m}$ include the density $\rho$, Young's modulus $E$, and Poisson's ratio $\nu$. For granular material, the material properties also include the friction angle $\theta$. The physics solver is built upon MPM~\cite{jiang2016material}, a hybrid Eulerian-Langrangian method. It simulates based on both particles and a spatial grid. As mentioned above, we densely sample particles inside the object in addition to the surface surfels. 
In each simulation time step, an MPM solver computes the momentum of each object particle (surfel) to update the dynamics attributes. In detail, the momentum of each particle is transferred to the grid within a particle-to-grid step, to further compute the terms like deformation gradient. These updates are back propagated into particles through a grid-to-particle process to update particle dynamics properties like position and velocity.

\myparagraph{Cloth and smoke.} 
We model smoke and cloth with only particles and employ the Position-Based Dynamics (PBD) solver~\citep{muller2007position} for these effects. 
The material properties $\mathbf{m}$ for cloth includes density $\rho$ and stretch/bending compliance $p$. The material properties for smoke includes $\rho$ and viscosity coefficient $\mu$.
We also densely sample particles inside smoke. Unlike MPM method, PBD method directly models the positions of each particle through a list of inequality and non-equality constraints. 
In each time step, PBD solver solves each constraint sequentially and directly update the particle's position which is then used to update the velocity. The constraints for smoke include incompressibility~\citep{macklin2013position}; the constraints for cloth include stretch and bending compliance. We refer the read to~\citet{bender2015position} for more information.

\subsection{Simulation Parameters}
\label{supp:params}

Different solvers rely on the different sets of physical parameters. Here we provide a table of all the parameters we set in the physical simulation process in Tab~\ref{tab:sup-params}, also with their default values. In simulation these parameters can be roughly estimated with a VLM and optional manual adjustment, as long as the simulation results are reasonable.

\subsection{Rendering Simulated Dynamics}

Upon physical simulation, we need to further map the simulated outputs to the Gaussian surfels for rendering the coarse dynamics. For rigid body objects, since each Gaussian surfel is initialized from one mesh vertex and the rigid body solver runs on the mesh representation, we can directly update each surfel's position with simulation results. For particle-based MPM and PBD solvers, during the initial sampling process, we record the mapping of each Gaussian surfel to the nearest $10$ sampled particles. At each simulation step, we use the average of position updates of these nearest particles to update the position of the corresponding Gaussian surfel.

\input{supp_text/params}

%% file: figTexts/staged_scenes.tex
\begin{figure}[t]
    \centering
    \includegraphics[width=\linewidth]{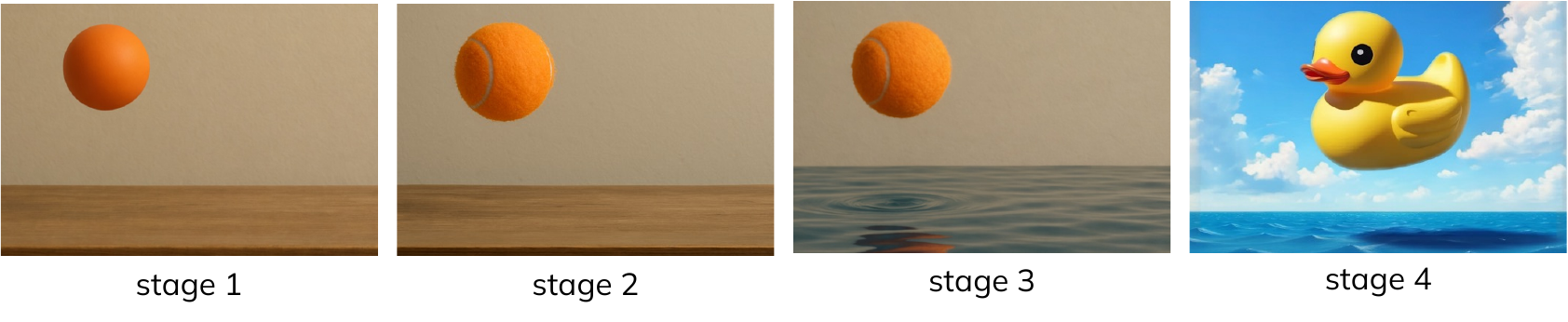}
    \caption{Scenes with increasing complexity. The first scene involves a rigid ball falling onto a rigid plane. The second replaces the rigid ball with a soft ball. The third scene replaces the rigid plane with water surface to include multi-physics. The fourth scene include an object with complex geometry.}
    \label{fig:staged_scenes}
\end{figure}

%% file: figTexts/staged_results.tex
\begin{figure}[t]
    \centering
    \includegraphics[width=\linewidth]{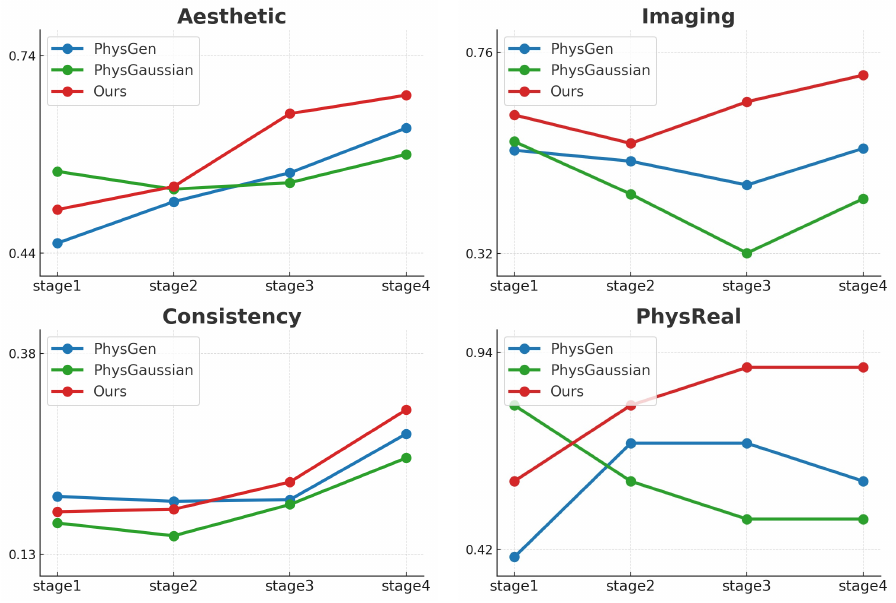}
    \caption{Quantitative results on the four stages (scenes with increasing complexity).}
    \label{fig:staged_results}
\end{figure}

%% file: tables/additional_ablation.tex
\begin{table}[t]
\centering
\scriptsize
\setlength{\tabcolsep}{1pt}
\renewcommand{\arraystretch}{1.2}
\begin{tabular}{lccccc}
\toprule
Methods & Imaging ($\uparrow$) & Aesthetic ($\uparrow$) & Motion ($\uparrow$) & Consistency ($\uparrow$) & PhysReal ($\uparrow$) \\
\midrule
\textbf{Ours} & $\mathbf{0.695}$ & \underline{$0.610$} & $\mathbf{0.995}$ & \underline{$0.217$} & $\mathbf{0.700}$ \\
Ours w/o RGB & $0.673$ & $0.601$ & $0.993$ & $0.212$ & $0.670$ \\
Ours w/o flow & $0.574$ & $0.587$ & \underline{$0.994$} & $0.213$ & $0.650$ \\
Coarse simulation & $0.552$ & $0.577$ & $\mathbf{0.995}$ & $0.197$ & $0.500$ \\
\midrule
$\scriptstyle s_1=19,s_2=16$ & $0.610$ & $0.571$ & \underline{$0.994$} & $0.215$ & $0.650$ \\
$\scriptstyle s_1=23,s_2=20$ & $0.683$ & $0.581$ & $\mathbf{0.995}$ & \underline{$0.217$} & \underline{$0.690$} \\
$\gamma = 0.3$ & $0.662$ & $0.571$ & \underline{$0.994$} & \underline{$0.217$} & $0.670$ \\
\bottomrule
\end{tabular}
\caption{Ablation study on conditioning signals and diffusion hyper-parameters.}
\vspace{-0.1cm}
\label{tab:additional_ablation}
\end{table}

%% file: supp_text/params.tex
\begin{table}[t]
\small
\begin{center}
\begin{tabular}{lc}
\toprule
 Parameter & Default Value \\ \midrule
\textbf{General simulation} &  \\
 Step time & $1e^{-2}$ \\
 Sub-steps number & $10$ \\
 Sampled particle size & $1e ^{-2}$ \\
 Gravity & $(0, 0, -9.8)$ \\

\midrule
 \textbf{Rigid body solver} &  \\
 friction coefficient & $0.1$ \\

\midrule
\textbf{MPM solver} & \\
Grid density & $128$ \\
Elastic material Young's modulus & $3e^5$ \\
Elastic material Poisson's ratio & $0.2$ \\

Liquid material Young's modulus & $1e^7$ \\
Liquid material Poisson's ratio & $0.2$ \\

Granular material Young's modulus & $1e^6$ \\
Granular material Poisson's ratio & $0.2$ \\
Granular material Friction angle & $45$ \\
\midrule

\textbf{PBD solver} & \\
Cloth material stretch compliance & $1e^{-7}$ \\
Cloth material bending compliance & $1e^{-5}$ \\
Smoke material viscosity coefficient & $0.1$ \\

\bottomrule
\end{tabular}
\end{center}
\caption{Simulation parameters and default values}\label{tab:sup-params}
\end{table}